\documentclass[pre,showpacs,amsfonts,amssymb,eqsecnum,%
twocolumn,floatfix,longbibliography,superscriptaddress]{revtex4-1}
\usepackage{amsmath}
\usepackage{graphicx}
\usepackage{bm}
\usepackage{xcolor}
\allowdisplaybreaks

\begin{document}
	
	\title{Continuous time random walks under Markovian resetting}
	
	\author{Vicen\c{c} M\'{e}ndez}
	\affiliation{Grup de F{\'i}sica Estad\'{i}stica.  Departament de F{\'i}sica.
		Facultat de Ci{\`e}ncies. Edifici Cc. Universitat Aut\`{o}noma de Barcelona,
		08193 Bellaterra (Barcelona) Spain}
	\author{Axel Mas\'{o}-Puigdellosas}
	\affiliation{Grup de F{\'i}sica Estad\'{i}stica.  Departament de F{\'i}sica.
		Facultat de Ci{\`e}ncies. Edifici Cc. Universitat Aut\`{o}noma de Barcelona,
		08193 Bellaterra (Barcelona) Spain}
		\author{Trifce Sandev}
	\affiliation{Research Center for Computer Science and Information Technologies, Macedonian Academy of Sciences and Arts, Bul. Krste Misirkov 2, 1000 Skopje, Macedonia} \affiliation{Institute of Physics \& Astronomy, University of Potsdam, D-14776 Potsdam-Golm, Germany} \affiliation{Institute of Physics, Faculty of Natural Sciences and Mathematics, Ss.~Cyril and Methodius University, Arhimedova 3, 1000 Skopje, Macedonia}
	\author{Daniel Campos}
	\affiliation{Grup de F{\'i}sica Estad\'{i}stica.  Departament de F{\'i}sica.
		Facultat de Ci{\`e}ncies. Edifici Cc. Universitat Aut\`{o}noma de Barcelona,
		08193 Bellaterra (Barcelona) Spain}

	\date{\today}
	
	\begin{abstract}
		We investigate the effects of markovian resseting events on continuous time random walks where the waiting times and the jump lengths are random variables distributed according to power law probability density functions. We prove the existence of a non-equilibrium stationary state and finite mean first arrival time. However, the existence of an optimum reset rate is conditioned to a specific relationship between the exponents of both power law tails. We also investigate the search efficiency by finding the optimal random walk which minimizes the mean first arrival time in terms of the reset rate, the distance of the initial position to the target and the characteristic transport exponents.
	\end{abstract}

\maketitle

\section{INTRODUCTION}

Diffusion with stochastic resetting was originally proposed some years ago \cite{EvMa11}. In this work a Brownian motion in an infinite medium is interrupted by resets events, which instantaneously returns the particle to the origin. The overall phenomena consists of two random processes independent of each other: the random motion of the particle and the resetting mechanism. Resets happen randomly in time according to a Poisson point process with definite intensity or constant rate. This resetting process is Markovian and the reset times are exponentially distributed. Actually, the distribution of reset times may be considered in general as a distribution with finite moments, where the first moment is precisely the inverse of the reset rate.      
The interest in this kind of problem essentially resides on two rather remarkable facts. Firstly, the verification that resetting stabilizes the random walk process, in the sense that a nonstationary process, as is the diffusion in an infinite medium, becomes stationary when it is affected by Markovian resetting mechanism.  Secondly, the fact that Markovian resetting may significantly reduce the mean first-passage time which, in turn, may yield a process with infinite first-passage time to reach the target in a finite time.
Many generalizations of the random walk, beyond the Brownian dynamics, have been proposed considering Markovian resets. For example, random walks in bounded domains \cite{Ch15,Pa19,DuLeLi19}, subdiffusion or superdiffusion in continuous space \cite{MaCaMe19,KuGu15,KuGu19}, superdiffusion in discrete space \cite{KuMaSa14}, Telegraphic random walks \cite{Ma19}, non-instantaneous returns \cite{Ma19b,PaKu19}, residence waiting times after resetting \cite{Ma19c} or diffusion in a potential landscape \cite{Pa15}. Other generalizations have been proposed by non-Markovian resetting events which may destroy the two interesting properties mentioned above. Indeed, power-law distribution of reset times have proved the non-existence of neither stationary state nor finite mean first passage time \cite{MaCaMe19}. Other works deal with spatial \cite{EvMa11p} or temporal \cite{PaKuEv16} dependence of the reset rate. 

Here we explore another generalization of diffusion with markovian
resetting, wherein the diffusive process between resets is substituted
with a process generated through the continuous time random walk (CTRW) scheme. In contrast to recent
works that combine Markovian resets with CTRW with finite-moment waiting times and power law distributions for jump lengths \cite{KuGu15}
or power law distributed waiting times and jump lengths with finite
moments \cite{KuGu19}, we assume that both waiting time and jump
length PDFs are distributed according to power laws. 

Unlike for superdiffusive transport, it is known that there is no optimum
reset rate for the MFAT of subdiffusive transport \cite{MaCaMe19}. Due to the long
tails of the waiting time PDF in subdiffusive transport the resetting
process always increases the MFAT and an optimum reset rate is never
found. However, the resetting process always helps the long tails
in superdiffusive transport to reduce the MFAT and get an optimum
reset rate.

Due to the trade-off between both heavy tails, it is not obvious the existence
of an optimum reset rate which minimizes the MFAT. We find the existence conditions and how the optimum reset rate and the optimum MFAT scale with the initial distance to the target. We find an exact analytic solution to the non-equilibrium stationary state (NEES) and the scaling property that its tail follows. The most efficient random walk (the random walk which minimizes the MFAT) is also studied in terms of the initial distance to the target.

The paper is organized as follows. In Section \ref{sec2} we present the CTRW formalism and derive a general expression for the survival probability of arriving to a given position. In Section \ref{sec3}, we consider resetting on the motion described by the CTRW model and calculate the NESS, the MFAT and study its optimality. We conclude the paper in Section \ref{sec4}.

\section{CTRW AND SURVIVAL PROBABILITY}
\label{sec2}

We consider a particle performing a random walk in continuous time.
The particle starts the motion from an initial position $x_{0}$ jumping
instantaneously to a new position where it waits for a time
before proceeding with the next jump. Jump lengths and waiting times
are independent identically distributed random variables distributed
according to the probability distribution functions (PDF) $\Phi(z)$
and $\varphi(t)$, respectively. The PDF $P(x,t)$ for the particle
position at time $t$ is given by the Montroll-Weiss equation \cite{MoWe65}
\begin{equation}
P(k,s)=\frac{p_{0}(k)\left[1-\varphi(s)\right]}{s\left[1-\varphi(s)\Phi(k)\right]}\label{eq:mw}
\end{equation}
where $P(k,s)=\int_{-\infty}^{\infty}e^{ikx}\int_{0}^{\infty}e^{-st}P(x,t)dxdt$
is the Fourier-Laplace transform of $P(x,t)$, $p_{0}(k)=\int_{-\infty}^{\infty}e^{ikx}P(x,t=0)dx$
is the Fourier transform of the initial condition, and $\varphi(s)$
and $\Phi(k)$ the Laplace ($\varphi(s)=\mathcal{L}\left[\varphi(t)\right](s)=\int_{0}^{\infty}e^{-st}\varphi(t)dt$)
and Fourier transforms ($\Phi(k)=\mathcal{F}[\Phi(x)](k)=\int_{-\infty}^{\infty}e^{ikx}\Phi(x)dx$)
of the waiting time and jump lengths PDFs respectively. Rearranging
Eq. (\ref{eq:mw}) in the form

\begin{equation}
s\left[\frac{1}{\varphi(s)}-\Phi(k)\right]P(k,s)=p_{0}(k)\left[\frac{1}{\varphi(s)}-1\right],\label{eq:mw2}
\end{equation}
after straightforward algebraic manipulations, we can rewrite Eq. (\ref{eq:mw2})
as
\begin{equation}
sP(k,s)-p_{0}(k)=K(s)\left[\Phi(k)-1\right]P(k,s),\label{eq:mw21}
\end{equation}
where we have defined the memory kernel
\begin{equation}
K(s)=\frac{s\varphi(s)}{1-\varphi(s)}.\label{eq:mk}
\end{equation}
By inverting Eq. (\ref{eq:mw21}) in Fourier-Laplace we obtain the
so called generalized CTRW master equation \cite{KeMo73}
\begin{equation}
\frac{\partial P}{\partial t}=\int_{0}^{t}K(t-t')\left[\int_{-\infty}^{\infty}P(x-z,t')\Phi(z)dz-P(x,t')\right]dt'.\label{eq:me}
\end{equation}
We are interested in power law PDFs for waiting time $\varphi(t)\sim t^{-(1+\gamma)}$ and jump lengths $\Phi(x)\sim |x|^{-(1+\alpha)}$ which, in the Laplace and Fourier spaces read

\begin{equation}
\varphi(s)=\frac{1}{1+(s\tau)^{\gamma}},\label{eq:pdf1}
\end{equation}
with $0<\gamma<1$ and
\begin{equation}
\Phi(k)\simeq1-\sigma^{\alpha}\left|k\right|^{\alpha}\label{eq:pdf2}
\end{equation}
with $1<\alpha<2$, respectively. Then $K(s)=s^{1-\gamma}/\tau^\gamma$ and Eq. (\ref{eq:mw2}) takes the form
\begin{equation}
\tau^{\gamma}s^{\gamma}P(k,s)-\tau^{\gamma}s^{\gamma-1}p_{0}(k)=-\sigma^{\alpha}\left|k\right|^{\alpha}P(k,s).\label{eq:mw3}
\end{equation}
By inverting in Fourier-Laplace we find the following fractional transport
equation (by defining the generalized diffusion coefficient $D=\sigma^\alpha/\tau^\gamma$)  
\begin{equation}
\frac{\partial^{\gamma}P}{\partial t^{\gamma}}=D\frac{\partial^{\alpha}P}{\partial\left|x\right|^{\alpha}}\label{eq:fd}
\end{equation}
where the Laplace transform of the Caputo fractional derivative
\[
\mathcal{L}\left[\frac{\partial^{\gamma}P(x,t)}{\partial t^{\gamma}}\right](s)=s^{\gamma}P(x,s)-s^{\gamma-1}P(x,t=0)
\]
and the Fourier transform of the Riesz fractional derivative
\[
\mathcal{F}\left[\frac{\partial^{\alpha}P(x,t)}{\partial\left|x\right|^{\alpha}}\right](k)=-\left|k\right|^{\alpha}P(k,s)
\]
 are introduced. The mean squared displacement that characterizes
the transport regime scales with time as $\left\langle x^{2}(t)\right\rangle \sim t^{2\gamma/\alpha}$
\cite{MeKl00} exhibiting normal diffusion ($2\gamma/\alpha=1$),
superdiffusion ($2\gamma/\alpha>1$) of subdiffusion ($2\gamma/\alpha<1$). 

To compute the MFAT under Markovian resetting it is necessary first
to find the survival probability $Q_{x_{0}}(t)$ up to time $t$ of
the transport process. In particular, $Q_{x_{0}}(t)$ is the probability
of not having reached the origin ($x=0$) in the first trip, which
ends at a random time $t'$, when particle starts the motion at position
$x_{0}$. Then the first arrival time probability density $f_{a}(x_{0},t)$ at origin
at a time $t$ for a particle that starts its motion
at position $x_{0}$ and the survival probability are related to each
other via
\begin{equation}
Q_{x_{0}}(t)=1-\int_{0}^{t}f_{a}(x_{0},t')dt'\label{eq:r1}
\end{equation}
or equivalently by $f_{a}(x_{0},t)=-\partial Q_{x_{0}}(t)/\partial t$.
We follow the method in \cite{ChMeGo03} since it has the advantage of being simpler than the traditional procedure of solving the equation for the propagator with an absorbing boundary at $x_0$. Furthermore, it has been shown to be valid even for L\'evy flights \cite{ChMeGo03}. We make use of the generalized master equation, which is written as a rate equation for the probability, with a $\delta$-sink
of a strength $f_{a}(x_{0},t)$
\begin{eqnarray}
& 
&\frac{\partial P(x,t)}{\partial t}\nonumber \\
&=&\int_{0}^{t}K(t-t')\left[\int_{-\infty}^{\infty}P(x-z,t')\Phi(z)dz-P(x,t')\right]dt' \nonumber \\ 
&-&f_{a}(x_{0},t)\delta(x)\label{eq:me1}
\end{eqnarray}
where $P(x,t)$ is now a non-normalized probability. We assume that
at $t=0$ the particle is placed at $x=x_{0}$, i.e., $P(x,t=0)=p_{0}(x)=\delta(x-x_{0}).$
Taking the Fourier-Laplace transform of Eq. (\ref{eq:me1}) and $p_{0}(k)=e^{ikx_{0}}$
one gets
\begin{eqnarray}
P(k,s)&=&\frac{e^{ikx_{0}}}{s+K(s)\left[1-\Phi(k)\right]}\nonumber \\ 
&-&f_{a}(x_{0},s)\frac{1}{s+K(s)\left[1-\Phi(k)\right]}.\label{eq:prop}
\end{eqnarray}
Since $f_{a}(x_{0},t)$ measures the first arrival time to the origin
we have taken into account the $\delta$-sink in Eq. (\ref{eq:me1}),
i.e., the origin is a perfectly absorbing boundary and $P(x=0,t)=0.$
We can exploit this property by arguing that the inverse Fourier transform
of $P(k,s)$ is
\begin{equation}
P(x,s)=\mathcal{F}^{-1}[P(k,s)](x)=\frac{1}{2\pi}\int_{-\infty}^{\infty}e^{-ikx}P(k,s)dk\label{eq:if}
\end{equation}
and taking $x=0$ we have 
\begin{equation}
\int_{-\infty}^{\infty}P(k,s)dk=0.\label{eq:cond}
\end{equation}
Integrating Eq. (\ref{eq:prop}) and taking into account Eq. (\ref{eq:cond})
we obtain 
\begin{eqnarray}
Q_{x_{0}}(s)&=&\frac{1}{s}\left[1-f_{a}(s)\right]\nonumber \\
&=&\frac{1}{s}\left[1-\frac{P(x=0,s;x_{0})}{P(x=0,s;0)}\right]\label{eq:spt}
\end{eqnarray}
once we have transformed Eq. (\ref{eq:r1}) by Laplace. Note that
the propagator $P(x,s;x_{0})$ is given, from Eq. (\ref{eq:mw3}),
by
\begin{equation}
P(x,s;x_{0})=\frac{1}{2\pi}\int_{-\infty}^{\infty}\frac{e^{-ik(x-x_{0})}}{s+K(s)\left[1-\Phi(k)\right]}dk.\label{eq:prop1}
\end{equation}
Considering (\ref{eq:pdf1}) and (\ref{eq:pdf2}) in (\ref{eq:prop1}), the propagator reads
\[
P(x,s;x_{0})=\frac{1}{\pi}\int_{0}^{\infty}\frac{\cos[k(x-x_{0})]dk}{s+s^{1-\gamma}\tau^{-\gamma}(\sigma k)^{\alpha}}
\]
which inserted in (\ref{eq:spt}) allows to find
\begin{equation}
Q_{x_{0}}(s)=
\frac{\alpha\sigma\sin\left(\frac{\pi}{\alpha}\right)}{\pi(s\tau)^{\gamma/\alpha}}\int_{0}^{\infty}\frac{1-\cos(kx_{0})}{s+s^{1-\gamma}\tau^{-\gamma}(\sigma k)^{\alpha}}dk
\label{qx0sr}
\end{equation}
where we made use of the result
\begin{equation}
\int_{0}^{\infty}\frac{dk}{(s\tau)^{\gamma}/\sigma^{\alpha}+k^{\alpha}}=\frac{\pi\sigma^{\alpha-1}}{\alpha\sin(\pi/\alpha)}(s\tau)^{\gamma(\frac{1}{\alpha}-1)}.
\label{deno}
\end{equation}
Inverting Eq. (\ref{qx0sr}) by Laplace  we find
\begin{eqnarray}
& &Q_{x_{0}}(t)=1-\frac{\alpha\sigma\sin\left(\frac{\pi}{\alpha}\right)}{\pi}(t/\tau)^{\gamma/\alpha}\times\nonumber\\
& &\int_{0}^{\infty}\cos(kx_{0})E_{\gamma,1+\frac{\gamma}{\alpha}}\left(-\sigma^{\alpha}k^{\alpha}\frac{t^{\gamma}}{\tau^{\gamma}}\right)dk=1-
\nonumber\\
& &{\sin\left(\frac{\pi}{\alpha}\right)H_{3,3}^{2,1}\left[\frac{|x_{0}|/\sigma}{(t/\tau)^{\gamma/\alpha}}\left|\begin{array}{ccc}
\left(\frac{\alpha-1}{\alpha},\frac{1}{\alpha}\right) & \left(1,\frac{\gamma}{\alpha}\right) & \left(\frac{1}{2},\frac{1}{2}\right)\\
(0,1) & \left(\frac{\alpha-1}{\alpha},\frac{1}{\alpha}\right) & \left(\frac{1}{2},\frac{1}{2}\right)
\end{array}\right.\right]}\nonumber\\
\label{qxo2}
\end{eqnarray}
where $E_{\alpha,\beta}(z)$ is the two-parametric Mittag-Leffler function \cite{Go14}
\begin{equation}
E_{a,b}(z)=\sum_{n=0}^{\infty}\frac{z^{n}}{\Gamma(an+b)}
    \label{mlf}
\end{equation}
and $H_{p,q}^{m,n}(z)$ is the Fox-H function \cite{MaSa10}. Known particular cases may be recovered from (\ref{qxo2}). For $\gamma=1$ and $\alpha =2$, the transport process corresponds to normal diffusion. Inserting these values in (\ref{qxo2}) and taking into account that in this case
$$
E_{1,3/2}(z)=\frac{e^{z}}{\sqrt{z}}\text{erf}(\sqrt{z})
$$
the integral in (\ref{qxo2}) can be straightforwardly computed to get
\begin{equation}
Q_{x_{0}}(t)=\text{erf}\left(\frac{\left|x_{0}\right|}{2\sigma}\sqrt{\frac{\tau}{t}}\right)=\text{erf}\left(\frac{\left|x_{0}\right|}{\sqrt{4Dt}}\right)
\label{qx0d}
\end{equation}
which corresponds to the known result in \cite{Re01} by considering $D=\sigma/\tau$. If we set $\gamma=1$ in (\ref{qxo2}) we recover the result found in \cite{ChMeGo03} for Lévy flights. 
The large time behaviour of $Q_{x_0}(t)$ can be found by using the series expansion of the Fox function in Eq.~(2.19) for $t\rightarrow\infty$ (small argument expansion of the Fox function), to get \cite{MaSa10}
\begin{eqnarray}
Q_{x_{0}}(t) &\approx& 
\frac{\alpha\sqrt{\pi}\,}{2^{\alpha}\,\Gamma(\frac{\alpha}{2})\,\Gamma(\frac{1+\alpha}{2})}\left(\frac{|x_0|}{\sigma}\right)^{\alpha-1}\frac{(t/\tau)^{-\frac{\gamma}{\alpha}(\alpha-1)}}{\Gamma(1-\gamma+\frac{\gamma}{\alpha})}\nonumber \\
&\sim& \frac{1}{t^{\frac{\gamma}{\alpha}(\alpha-1)}}.
\label{qx0d2}
\end{eqnarray}
Introducing $\alpha =2$ in (\ref{qx0d2}) with $0<\gamma<1$ (subdiffusion) and $\gamma =1$ with $1<\alpha<2$ (Lévy flights) we recover the known results $Q_{x_{0}}(t)\sim t^{-\gamma /2}$ \cite{RaDi00} and $Q_{x_{0}}(t)\sim t^{-1+1/\alpha}$ \cite{ChMeGo03}, respectively. From Eq. (\ref{eq:r1}) the long term behavior of the first arrival time PDF in these two cases is $f_a(x_{0},t)\sim t^{-\gamma /2-1}$  and $f_a(x_{0},t)\sim t^{-2+1/\alpha}$, respectively.
For computing purposes is interesting to express the Fox function in Eq. (\ref{qxo2}) as a power series. By using Ref. \cite{MaSa10} we find, after some simplifications,
\begin{eqnarray}
 Q_{x_{0}}(t)&=&\sin\left(\frac{\pi}{\alpha}\right)\nonumber \\
&\times&\left[\sum_{n=1}^{\infty}\frac{(-1)^{n+1}\sin\left(\frac{\pi(1+n)}{2}\right)\left(\frac{|x_{0}|}{\sigma}\right)^{n}}{n!\sin\left(\frac{\pi(1+n)}{\alpha}\right)\Gamma\left(1-\frac{\gamma n}{\alpha}\right)(\frac{t}{\tau})^{\gamma n/\alpha}}\right.\nonumber \\
&+&\left.\frac{\alpha}{\sqrt{\pi}}\sum_{n=1}^{\infty}\frac{2^{-\alpha n}(-1)^{n}\Gamma\left(\frac{1-n\alpha}{2}\right)\left(\frac{\left|x_{0}\right|}{\sigma}\right)^{n\alpha-1}}{\Gamma\left(\frac{n\alpha}{2}\right)\Gamma\left(1+\frac{\gamma}{\alpha}-\gamma n\right)(\frac{t}{\tau})^{\frac{\gamma}{\alpha}(n\alpha-1)}}\right]\nonumber\\
\label{qx03}
\end{eqnarray}

To our knowledge, this is the first time that the general expressions (\ref{qxo2}) and (\ref{qx03}) and the scaling (\ref{qx0d2}) have been reported in such a general form.

\section{CTRW AND MARKOVIAN RESETTING}
\label{sec3}

When resetting is taken into account the particle moves according
to the CTRW propagator (\ref{eq:mw}) during a period called reset
time and then the particle jumps instantaneously to the initial position $x=x_0$ to start its motion again. Then
the incorporation of the resetting process to the transport process
results in a sequence of transport periods and instantaneous resets
to $x_0$. The time spent between two consecutive resets is the
reset time and is a random variable distributed according to the PDF
$\varphi_{R}(t)$. A general formulation for the combination of a
general transport with resetting has been recently studied \cite{MaCaMe19}
where the NESS and the MFAT have been obtained for any transport propagator and for any resetting PDF. 

The propagator $\rho(x,t;x_0)$ of the joint process (transport
and resetting to the initial position $x_0$) is given by the renewal equation \cite{MaCaMe19}
\begin{equation}
\rho(x,t;x_0)=\varphi_{R}^{*}(t)P(x,t;x_0)+\int_{0}^{t}\varphi_{R}(t')\rho(x,t-t';x_0)dt'\label{eq:re1}
\end{equation}
where $\varphi_{R}^{*}(t)=\int_{t}^{\infty}\varphi_{R}(t')dt'$ is
the probability of the first reset happening after time $t$. In the
Laplace space this equation has the form 

\begin{equation}
\rho(x,s;x_0)=\frac{\mathcal{L}\left[\varphi_{R}^{*}(t)P(x,t;x_0)\right](s)}{1-\varphi_R(s)}.\label{re2}
\end{equation}

The survival probability $S_{x_{0}}(t)$ of the joint process
(transport and resetting) is given by the renewal equation 

\begin{equation}
S_{x_{0}}(t)=\varphi_{R}^{*}(t)Q_{x_{0}}(t)+\int_{0}^{t}\varphi_{R}(t')Q_{x_{0}}(t')S_{x_{0}}(t-t')dt'\label{eq:sp2}
\end{equation}
where $Q_{x_{0}}(t)$ is the survival probability of the transport
process and is given by Eq. (\ref{eq:spt}). Eq. (\ref{eq:sp2}) can
be solved in the Laplace space
\begin{equation}
S_{x_{0}}(s)=\frac{\mathcal{L}\left[\varphi_{R}^{*}(t)Q_{x_{0}}(t)\right](s)}{1-\mathcal{L}\left[\varphi_{R}(t)Q_{x_{0}}(t)\right]}.\label{eq:sp3}
\end{equation}
The total (transport and resetting) first arrival time probability
density at origin $F_{a}(x_{0},t)$ is obtained in terms of $S_{x_{0}}(t)$:
$F_{a}(x_{0},t)=-\partial S_{x_{0}}(t)/\partial t$. Then the
MFAT is given by

\begin{eqnarray}
T(x_{0})&=&\int_{0}^{\infty}tF_{a}(x_{0},t)dt=\lim_{s\rightarrow0}S_{x_{0}}(s)\nonumber \\ 
&=&\frac{\int_{0}^{\infty}\varphi_{R}^{*}(t)Q_{x_{0}}(t)dt}{1-\int_{0}^{\infty}\varphi_{R}(t)Q_{x_{0}}(t)dt}.\label{eq:mfat1}
\end{eqnarray}

Eq. (\ref{eq:mfat1}) provides then the mean time that a particle needs to arrive for the first time at the origin $x=0$, where a target is located. Likewise, $|x_0|$ is the initial distance between the particle and the target. The particle, which starts its motion at $x=x_0$ (whose survival probability is $Q_{x_{0}}(t)$) is reset to $x=x_0$ after a random time distributed according to the PDF $\varphi_{R}(t)$. Then, the MFAT in Eq. (\ref{eq:mfat1}) is a measure of the search efficiency when the particle movement is described by a CTRW under a resetting mechanism. 

To obtain specific results for the NESS and the MFAT we need
to consider particular expressions for $\varphi_{R}(t)$. Thorough
this work we consider that resetting is a Markovian process, i.e.,
the reset rate is exponentially distributed
\begin{equation}
\varphi_{R}(t)=re^{-rt}.\label{eq:exp}
\end{equation}
Below we apply these results to the CTRW propagator in Eq. (\ref{eq:mw}). Considering Eqs. (\ref{eq:mfat1},\ref{eq:exp}) it is found

\begin{equation}
T(x_{0})=\frac{Q_{x_0}(s=r)}{1-rQ_{x_0}(s=r)},
\label{mfat32}
\end{equation}
where $Q_{x_0}(s=r)$ is found from Eq. (\ref{qx0sr}).

\subsection{NESS}

\begin{figure}[htbp]
	\includegraphics[width=\hsize]{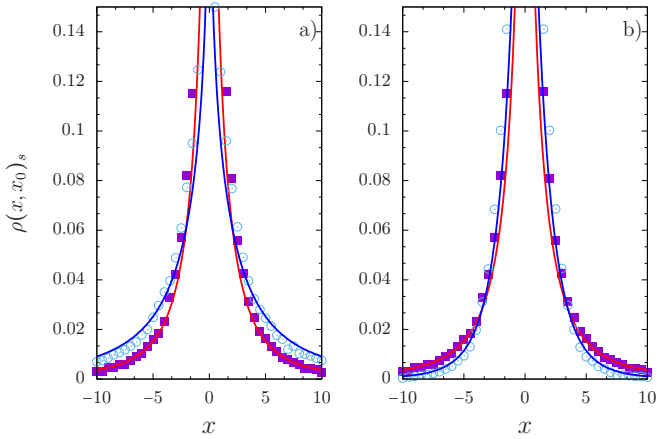}
	\caption{NEES for different values of $\alpha$ and $\gamma$ with $\sigma =r=1$, $\tau=0.1$, $x_0=0$. In panel a) we fix $\alpha=1.25$. The cases $\gamma = 0.25$ and $\gamma=0.75$ are drawn in red and blue respectively. In panel b) we fix $\gamma=0.25$. The cases $\alpha = 1.25$ and $\alpha=1.75$ are drawn in red and blue respectively. Analytical solution is computed from Eq. (\ref{nees}) and shown with solid curves while symbols correspond to numerical simulations.}
	\label{fig:f1}
\end{figure}

Applying the Fourier transform to Eq.\eqref{re2} and introducing the propagator in \eqref{eq:prop1} and (\ref{eq:exp}), one can obtain the propagator of the joint process in the Fourier-Laplace space to be

\begin{equation}
\rho(k,s;x_0)=\frac{1}{2\pi s} \int_{-\infty}^{\infty} \frac{(r+s)e^{-ik(x-x_0)}dk}{s+r+K(s+r)[1-\Phi(k)]}.
\label{prop2}
\end{equation}
To obtain the stationary solution of the joint process we take the limit $s\rightarrow0$ to Eq. (\ref{prop2}) and inserting Eqs. (\ref{eq:pdf1},\ref{eq:pdf2})
we finally get the exact expression for the NEES:
\begin{eqnarray}
& &\rho_{s}(x,x_{0})=\frac{1}{\pi}\int_{0}^{\infty}\frac{\cos\left[k(x-x_{0})\right]}{1+(r\tau)^{-\gamma}\sigma^{\alpha}k^{\alpha}}dk \nonumber \\ 
&=&\frac{1}{\pi}\int_{0}^{\infty}\cos\left[k(x-x_{0})\right]H_{1,1}^{1,1}\left[\frac{\sigma^{\alpha}k^{\alpha}}{(r\tau)^{\gamma}}\left|\begin{array}{c}
(0,1)\\
(0,1)
\end{array}\right.\right]dk\nonumber \\ 
&=&\frac{1}{\left|x-x_{0}\right|}H_{2,3}^{2,1}\left[\frac{(r\tau)^{\gamma}\left|x-x_{0}\right|^{\alpha}}{\sigma^{\alpha}}\left|\begin{array}{ccc}
(1,1) & (1,\frac{\alpha}{2})\\
(1,\alpha) & (1,1) & (1,\frac{\alpha}{2})
\end{array}\right.\right].\nonumber \\
\label{nees}
\end{eqnarray}
In Figure \ref{fig:f1} we show a comparison of the NEES obtained in (\ref{nees}) with numerical simulations sampling dispersal distances from a Lévy PDF ($\Phi(k)=e^{-\sigma^\alpha |k|^\alpha}$). It is noticeable that to get analytic results we have considered actually an approximation to the Lévy PDF for large dispersal distances in Eq. (\ref{eq:pdf2}) ($\Phi(k)=e^{-\sigma^\alpha |k|^\alpha}\approx 1-\sigma^\alpha |k|^\alpha$). For this reason the agreement between numerical and theoretical results fails close to $x=0$ (not shown). In panel a) we show the case $\gamma=0.25$ (in red) and $\gamma=0.75$ (in blue) for fixed $\alpha=1.25$. We see how the exponent $\gamma$ modify the shape of the NEES. In panel b) we consider $\gamma=0.25$ and the cases $\alpha=1.25$ (in red) and $\alpha=1.75$ (in blue). The tail decays faster with $x$ as $\alpha$ increases, as we also show in turn analytically.  The Fox function admits series expansion for $\left|x-x_{0}\right|\gg\sigma$
 \cite{MaSa10}. Taking the lowest order we obtain the following scaling
for the tail of the NEES:
\begin{equation}
\rho_{s}(x,x_{0})\sim\frac{\sigma^{\alpha+1}}{\left|x-x_{0}\right|^{\alpha+1}}.\label{eq:tail}
\end{equation}
 It is interesting to note that this scaling behavior is not affected
by the waiting time PDF tail, i.e., the tail of the NEES is controlled
by the jump length PDF only. 

\subsection{MFAT}

The MFAT is found by introducing Eq. (\ref{qx0sr}) into Eq. (\ref{mfat32}) to get, after some manipulations
\begin{equation}
T(x_{0})=\frac{1}{r}\left[\frac{\pi(r\tau)^{\gamma/\alpha}}{\alpha\sigma\sin\left(\frac{\pi}{\alpha}\right)I(x_{0},r)}-1\right]
\label{eq:mfat3}
\end{equation}
where

\begin{equation}
I(x_0,r)=\int_{0}^{\infty}\frac{\cos(kx_{0})}{1+(r\tau)^{-\gamma}\sigma^{\alpha}k^{\alpha}}dk
\label{I}
\end{equation}
is defined.

This expression may be evaluated analytically by using Ref. \cite{MaSa10}
\begin{eqnarray}
&&I(x_0,r)=\int_{0}^{\infty}\cos(kx_{0})H_{1,1}^{1,1}\left[\frac{\sigma^{\alpha}k^{\alpha}}{(r\tau)^{\gamma}}\left|\begin{array}{c}
(0,1)\\
(0,1)
\end{array}\right.\right]dk\nonumber \\
&=&\frac{\pi}{\alpha\left| x_{0}\right|}H_{2,3}^{2,1}\left[\frac{(r\tau)^{\frac{\gamma}{\alpha}}\left|x_{0}\right|}{\sigma}\left|\begin{array}{ccc}
(1,\frac{1}{\alpha}) & (1,\frac{1}{2})\nonumber \\
(1,1) & (1,\frac{1}{\alpha}) & (1,\frac{1}{2})
\end{array}\right.\right].\nonumber \\
\label{eq:pxx0}
\end{eqnarray}
From (\ref{eq:mfat3}) and (\ref{eq:pxx0}) the MFAT takes finally the form 
\begin{equation}
T(x_{0})=\frac{1}{r}\left\{ \frac{\frac{\left|x_{0}\right|}{\sigma}(r\tau)^{\frac{\gamma}{\alpha}}[\sin(\frac{\pi}{\alpha})]^{-1}}{H_{2,3}^{2,1}\left[\frac{(r\tau)^{\frac{\gamma}{\alpha}}\left|x_{0}\right|}{\sigma}\left|\begin{array}{ccc}
(1,\frac{1}{\alpha}) & (1,\frac{1}{2})\\
(1,1) & (1,\frac{1}{\alpha}) & (1,\frac{1}{2})
\end{array}\right.\right]}-1\right\} .\label{eq:mfatex}
\end{equation}
Since there is no additional convergence restriction in calculating
the integral in Eq. (\ref{eq:pxx0}) we can assert
that the MFAT is finite for any set of parameters $\gamma$ and
$\alpha$, i.e., for any specific transport process. Eq. (\ref{eq:mfatex})  holds for any Markovian resetting (i.e, for a resetting time PDF with finite moments, since $r$ is nothing but the inverse of the first moment) with a general CTRW, since the expressions for the PDFs in Eqs.~(\ref{eq:pdf1}) and (\ref{eq:pdf2}) may lead to subdiffusive, superdiffusive or normal transport \cite{MeKl00}.
Making use of the power expansions of the Fox functions \cite{MaSa10} we can approximate (\ref{eq:mfat3}) in the limits $r\tau\ll1$ (small reset rate) and $r\tau\gg1$ (large reset rate).  For $r\tau\ll1$ the leading terms in Eq.~(\ref{eq:pxx0}) are
\begin{eqnarray}
I(x_0,r)&\simeq&\Gamma(1-1/\alpha)\Gamma(1/\alpha)\frac{(r\,\tau)^{\gamma/\alpha}}{\alpha\sigma}\nonumber\\&+&\frac{\pi\Gamma(1-\alpha)}{\Gamma(\alpha/2)\Gamma(1-\alpha/2)}\frac{(r\tau)^{\gamma}|x_0|^{\alpha-1}}{\sigma^{\alpha}}
\end{eqnarray}
which can be inserted in (\ref{eq:mfat3}) to get
\begin{equation}
T(x_{0})\approx\frac{\tau\alpha\Gamma(2-\alpha)\sin\left(\frac{\pi}{\alpha}\right)\sin\left(\frac{\pi\alpha}{2}\right)}{\pi(\alpha-1)(r\tau)^{\gamma\left(\frac{1}{\alpha}-1\right)+1}}\left(\frac{|x_{0}|}{\sigma}\right)^{\alpha-1}.
\label{Tx01}
\end{equation}
Analogously, for $r\tau\gg1$ the leading terms in Eq.~(\ref{eq:pxx0}) are
\begin{eqnarray}
I(x_0,r)&\simeq&\frac{\Gamma(1+\alpha)\sin(\pi\alpha/2)\,\sigma^{\alpha}}{(r\,\tau)^{\gamma}|x_0|^{\alpha+1}}
\end{eqnarray}
which leads us to
\begin{equation}
T(x_{0})\approx\frac{\tau\pi(r\tau)^{\gamma\left(1+\frac{1}{\alpha}\right)-1}}{\alpha\sin\left(\frac{\pi}{\alpha}\right)\sin\left(\frac{\pi\alpha}{2}\right)\Gamma (1+\alpha)}\left(\frac{|x_{0}|}{\sigma}\right)^{1+\alpha}.
\label{Tx02}
\end{equation}

\subsection{Optimal reset rate}
The question
is now to find the condition whether there is an optimal MFAT
and more specifically, what is the set of values of parameters $(\gamma,\alpha)$
for which there is a reset rate that optimizes the MFAT. To
this end we analyze the behavior of $T(x_{0})$ in the limits $r\rightarrow0^{+}$
and $r\rightarrow\infty.$ For fixed $x_{0}$, the MFAT in the limit $r\rightarrow0^{+}$ is given in Eq. (\ref{Tx01}), that is, 
$T(x_{0})\sim r^{\gamma\left(1-\frac{1}{\alpha}\right)-1}$. Since $\gamma \in (0,1)$ and $\alpha \in (1,2)$ the exponent is such that $\gamma\left(1-1/\alpha\right)<1$, i.e., it is always negative.
In consequence, $T(x_{0})\rightarrow\infty$
as $r\rightarrow0 ^{+}$. In addition, $T(x_{0})$ is a decreasing function with $r$ when $r$ takes small values. On the other hand, when $r$ takes large values the MFAT is given in Eq. (\ref{Tx02}), i.e, $T(x_{0})\sim r^{\gamma/\alpha+\gamma-1}$, which is an increasing function of $r$ only if $\gamma > \gamma_c$ with

\begin{equation}
\gamma_{c}=\frac{\alpha}{\alpha+1}.\label{eq:gc}
\end{equation}
Moreover, $T(x_{0})$ is a monotonically decreasing function of $r$ when $r$ is large if $\gamma <\gamma_c$, which means that even if there is a local minimum for a given $r$, since it is still decreasing with $r$, then the minimum MFAT is $T(x_{0})=0$. In consequence, there exists a nonzero minimum MFAT if $\gamma \geq \gamma_c$. Note that when $\gamma = \gamma_c$ two different situations could happen: the minimum is either attained at $r\rightarrow\infty$ or for a specific value of $r$. Since  $T(x_{0})$ is a decreasing function of $r$ for small $r$ and it tends to a constant value as $r\rightarrow\infty$, if $\partial T(x_{0})/\partial r <0$ the minimum MFAT is attained asymptotically at $r\rightarrow\infty$. Contrarily, if $\partial T(x_{0})/\partial r >0$ for large $r$ the minimum MFAT is attained at a given $r$. To uncover which is the actual situation we take $\gamma=\gamma_c$ in (\ref{Tx02}) to get

$$
T(x_{0})\approx\frac{\pi\tau(\left|x_{0}\right|/\sigma)^{1+\alpha}}{\alpha\sin(\pi/\alpha)\Gamma(1+\alpha)\sin(\pi\alpha/2)}-\frac{1}{r}
$$
It is easy to check that $\partial T(x_{0})/\partial r\approx 1/r^2>0$, so that a local minimum is attained when $\gamma=\gamma_c$. Therefore, we can conclude that there exists an optimum reset rate $r$ which minimizes $T(x_{0})$ if and only if
$$
\gamma \geq \gamma_c.
$$
\begin{figure}[htbp]
	\includegraphics[width=\hsize]{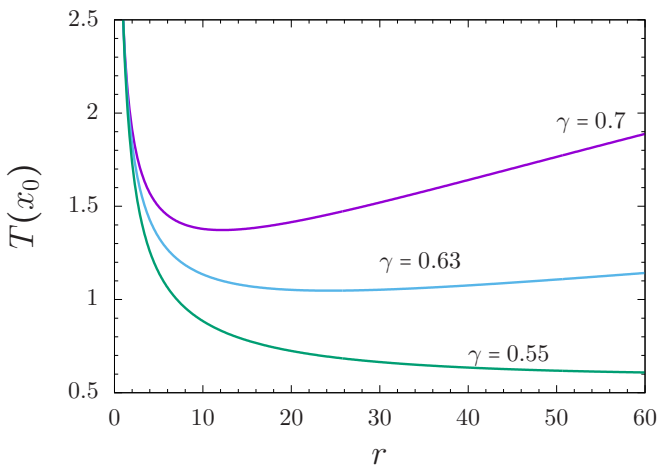}
	\caption{MFAT computed from Eq. (\ref{eq:mfatex}) for different values of $\gamma$. Parameters values are $x_0=\tau=\sigma=1$ and $\alpha=1.75$. $\gamma_c=0.636$. }
	\label{fig:f2}
\end{figure}

This is one of the main results of this work. In Figure 2 we plot the MFAT given by (\ref{eq:mfatex}) for different values of $\gamma$ above, below and at $\gamma=\gamma_c$. As can be seen for $\gamma=\gamma_c$ there is a local minimum for the MFAT. 
The optimal reset rate is obtained by solving numerically the equation $\partial T(x_{0})/\partial r =0$. Taking the derivative of (\ref{eq:mfatex}) and using the properties of the derivatives of the Fox functions \cite{MaSa10} one finds, after some simplifications, that the optimal reset rate $r^*$ is given by

\begin{equation}
r^* = \frac{1}{\tau}\left[\frac{\sigma}{|x_0|}z(\alpha,\gamma)\right]^{\alpha/\gamma}
\label{ropt}
\end{equation}
where $z=z(\alpha,\gamma)$
is the solution to the equation 
\begin{eqnarray}
1&-&\frac{\gamma}{\alpha}+\frac{H_{2,3}^{2,1}\left[z\left|\begin{array}{ccc}
(0,\frac{1}{\alpha}) & (1,\frac{1}{2})\nonumber\\
(1,1) & (1,\frac{1}{\alpha}) & (1,\frac{1}{2})
\end{array}\right.\right]}{H_{2,3}^{2,1}\left[z\left|\begin{array}{ccc}
(1,\frac{1}{\alpha}) & (1,\frac{1}{2})\nonumber\\
(1,1) & (1,\frac{1}{\alpha}) & (1,\frac{1}{2})
\end{array}\right.\right]}\nonumber \\
&=&\frac{\sin(\pi/\alpha)}{z}H_{2,3}^{2,1}\left[z\left|\begin{array}{ccc}
(1,\frac{1}{\alpha}) & (1,\frac{1}{2})\nonumber\\
(1,1) & (1,\frac{1}{\alpha}) & (1,\frac{1}{2})
\end{array}\right.\right]\\
\label{eqte}
\end{eqnarray}

From (\ref{ropt}) it is found the scaling dependence $r^*\sim |x_0|^{-\alpha/\gamma}$ which generalizes recent results obtained for any $\alpha$ and $\gamma =1$ \cite{KuGu15}.
If (\ref{ropt}) is introduced in Eq. (\ref{eq:mfatex}) then the optimum MFAT obeys the scaling relation
\begin{equation}
T^*(x_0) = \left(\frac{|x_0|}{\sigma}\right)^{\alpha/\gamma}T^*(\sigma) 
\label{topt1}
\end{equation}
which is again a generalization of the scaling found in \cite{KuGu15}. Here, $T^*(x_0)$ grows with $x_0$ faster than for $\gamma =1$ due to the effect of the heavy tailed waiting time PDF which slows down the search process of reaching the target at the origin.

\subsection{Optimal random walk}

Analogously,
fixing the value of the reset rate $r$, the optimal random walk (characterized by
the values of the exponents $(\gamma,\alpha)$ ) which minimizes the MFAT, depends on the distance
between the initial position of the random walker $x_{0}$ and the target point $x=0$. 
Let us consider two limiting situations: $x_{0}$ small and large. 
We assume $x_0>0$ for simplicity, otherwise we should replace $x_0$ by $|x_0|$ form now on.
As can be seen from (\ref{eq:mfatex}) the limit for small $x_{0}$ is equivalent to consider $r\tau\ll 1$. So that, when $x_{0}$ is small the MFAT is given by (\ref{Tx01})

As can be checked numerically, the minimum value of $T(x_{0})$ is
attained when $\alpha=2$ for the other parameters fixed, so the Brownian
motion is the most effective random walk when the particle starts the motion close to the resetting point. The optimal value is
 $$T^{*}(x_{0})\approx\tau(|x_{0}|/\sigma)(r\tau)^{-1+\gamma/2}$$ 
and is decreasing with $r$. This means that when the initial position of the particle is close to the target, the search process is more efficient when the particle motion has a short-tailed jump length distribution, being Brownian when $\gamma=1$.

Analogously, the limit for large $x_{0}$ is equivalent to consider $r\tau\gg 1$ and
the MFAT is given by (\ref{Tx02}). For fixed $r$, $x_0$ and $\gamma$, this expression has a minimum for a given value of the exponent $\alpha$ within the interval $(1,2)$ which means that L\'evy-like jumps are optimal when the initial position of the particle $x_0$ is far from the target position. Although the prefactors depend explicitly on $\gamma$, the scalings in (\ref{Tx01}) and (\ref{Tx02}) of $T(x_0)$ on $x_0$ only depend on the exponent $\alpha$ as in \cite{KuGu19}. 

\begin{figure}[htbp]
	\includegraphics[width=\hsize]{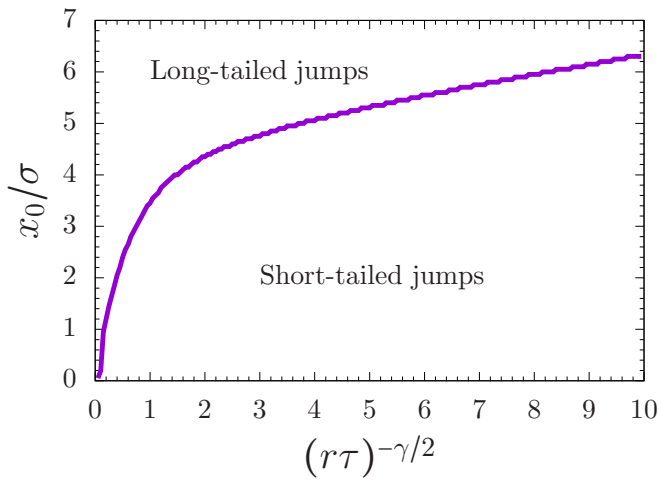}
	\caption{Optimality regions in the parameter space. We take $\sigma=\tau=1$ and $\gamma=0.5$ and vary  $x_0$ and $r$.}
	\label{fig:f3}
\end{figure}

However, as can be shown numerically, the MFAT given in (\ref{Tx02}) has a minimum for a given value of $\alpha$ which depends on $x_0$, $r$ and $\gamma$. Let $\alpha_{opt}(r,x_0,\gamma)$ denote the value of $\alpha$ at which the MFAT is minimum. If the minimum MFAT is attained at $\alpha_{opt}\in [1,2)$ then long-tailed jump distributions are optimal for the search strategy but if it is attained at $\alpha_{opt}=2$ then short-tailed jump distributions are the optimal. 
In order to gain a deeper understanding we can draw a phase diagram for the optimality regions by inspecting the minimum MFAT numerically. More specifically, we have computed Eq.\eqref{eq:mfat1} for $\sigma=\tau=1$ and $\gamma=0.5$ and different values of $x_0$ and $r$, as shown in Fig.\ref{fig:f3}. For each set of parameters, we have determined whether the MFAT has a minimum between  at $\alpha_{opt}\in [1,2)$ or at $\alpha_{opt}=2$ and have depicted the frontier. 

For parameter values below the critical curve, the MFAT has a minimum at $\alpha_{opt}=2$, i.e. short-tailed jump distributions are optimal. For parameter values above the critical curve the minimum MFAT is attained for a value of $\alpha$ between 1 and 2 and, therefore, long-tailed jump distributions attain the optimal MFAT. This, in the particular $\gamma=1$ case, shows that, on one hand, Brownian motion would be the optimal strategy to find a target near the origin. On the other hand, when the target is far from the origin, Levy flights become the optimal strategy to find it.

\section{Conclusions}
\label{sec4}
In this paper we considered a random walk with waiting times and jump lenghts distributed according to power laws and interrupted by a markovian resetting process. The walker starts the motion at point $x_0$ and a target is assumed to be located at $x=0$. First we found the exact solution and the scaling in the long time limit for the survival probability of not having reached the target in the first trip in absence of resetting. Second, we obtained exact analytic solutions to the NESS and the MFAT in presence of markovian resetting. Due to the opposite effect of the heavy tails of the waiting times and jump lenghts PDfs, there is a critical value of the waiting time exponent $\gamma$ for the existence of an optimum MFAT. We have obtained the critical exponent $\gamma$ and have found the parameters regions for optimal MFAT. We have also determined which type of motion strategy (diffusion or Levy flight) is optimal depending on the position of the target. We have found that for any reset rate $r$, one always finds a transition between optimal Brownian motion (small $x_0$) and optimal Levy flight (large $x_0$). This transition is also shown to exist for any choice of the waiting time distribution of the form in Eq.\eqref{eq:pdf1}, including long-tailed waiting times. Therefore, depending on the environmental ($x_0,\ r$) and the internal ($\gamma,\ \tau,\ \sigma$) parameters of the walker, any type of motion from subdiffusion to superdiffusion can be optimal to find the target.

\section*{Acknowledgments}
This research was partially supported by Grant No. CGL2016-78156-C2-2-R (V.M., D.C., A.M.). TS was supported by the Alexander von Humboldt Foundation.

\bibliography{ctrwresets}

\end{document}